\documentclass[showpacs,showkeys,preprintnumbers,prfluid,amsmath,amssymb,superscriptaddress]{revtex4-2}
\usepackage{setspace}
\usepackage{threeparttable} 
            
\usepackage{lineno,hyperref}
\usepackage[normalem]{ulem}
\usepackage{graphics}
\usepackage{amsmath,bm}
\usepackage{amssymb}
\usepackage{graphicx}
\usepackage{bm}
\usepackage{amsfonts}
\usepackage{color}
\usepackage{epstopdf}
\usepackage{array}
\usepackage{caption}
\usepackage{hyperref}
\usepackage{epic,eepic}
\usepackage{epsfig}
\usepackage{pifont}
\usepackage{xspace}
\usepackage[label font=bf,labelformat=simple]{subfig}
\usepackage{mathrsfs}
\usepackage{url}
\usepackage{booktabs}
\usepackage{float}

\usepackage[toc,page]{appendix}
\usepackage[capitalize]{cleveref}
\usepackage{mathtools}
\usepackage{gensymb}

\usepackage{algorithm}
\usepackage{algorithmic}

\newcommand{\be}{\begin{equation}}
\newcommand{\ee}{\end{equation}}
\newcommand{\ben}{\begin{equation*}}
\newcommand{\een}{\end{equation*}}

\newcommand{\bv}[1]{\mathbf{#1}}

\newcommand{\vv}{\bv{u}}

\newcommand{\delF}[1]{\textcolor{blue}{}}

\modulolinenumbers[5]

\begin{document}


\title{Near-wall approximations to speed up simulations for atmosphere boundary layers in the presence of forests using lattice Boltzmann method on GPU}


\author{Xinyuan Shao}
\affiliation{Department of Mechanics and Maritime Sciences, Division of Fluid Dynamics, Chalmers University of Technology, 41296, G{\"o}teborg, Sweden}

\author{Marta Camps Santasmasas}
\affiliation{School of Mechanical, Aerospace and Civil Engineering, The University of Manchester, Sackville Street, M1 3BB, United Kingdom}

\author{Xiao Xue}
\email{Corresponding author\\His email: xiaox@chalmers.se}
\affiliation{Department of Mechanics and Maritime Sciences, Division of Fluid Dynamics, Chalmers University of Technology, 41296, G{\"o}teborg, Sweden}

\author{Jiqiang Niu}
\affiliation{School of Mechanical Engineering, Southwest Jiaotong University, Chengdu, 610031, Sichuan, China}
\affiliation{State Key Laboratory of Automotive Simulation and Control, Jilin University, Changchun, 130025, Jilin, China}

\author{Lars Davidson}
\affiliation{Department of Mechanics and Maritime Sciences, Division of Fluid Dynamics, Chalmers University of Technology, 41296, G{\"o}teborg, Sweden}

\author{Alistair J. Revell}
\affiliation{School of Mechanical, Aerospace and Civil Engineering, The University of Manchester, Sackville Street, M1 3BB, United Kingdom}

\author{Hua-Dong Yao}
\affiliation{Department of Mechanics and Maritime Sciences, Division of Fluid Dynamics, Chalmers University of Technology, 41296, G{\"o}teborg, Sweden}

\begin{abstract}
Forests play an important role in influencing the wind resource in atmospheric boundary layers and the fatigue life of wind turbines. Due to turbulence, a difficulty in the simulation of the forest effects is that flow statistical and fluctuating content should be accurately resolved using a turbulence-resolved CFD method, which requires a large amount of computing time and resources. In this paper, we demonstrate a fast but accurate simulation platform that uses a lattice Boltzmann method with large eddy simulation on Graphic Processing Units (GPU). The simulation tool is the open-source program, GASCANS, developed at the University of Manchester. The simulation platform is validated based on canonical wall-bounded turbulent flows. A forest is modelled in the form of body forces injected near the wall. Since a uniform cell size is applied throughout the computational domain, the averaged first-layer cell height over the wall reaches to $\langle \Delta y^+\rangle = 165$. 
Simulation results agree well with previous experiments and numerical data obtained from finite volume methods. We demonstrate that good results are possible without the use of a wall-function, since the forest forces overwhelm wall friction. This is shown to hold as long as the forest region is resolved with several cells. 
In addition to the GPU speedup, the approximations also significantly benefit the computation efficiency. (This paper is currently submitted to peer-review journal) 
\end{abstract}

\keywords{Wind energy, Atmospheric boundary layers, Forest modelling, Lattice Boltzmann method, Large-eddy simulation, GPU}
\maketitle


\section{Introduction}

When wind farms are located within forest terrain, the presence of trees are known to generate strong atmospheric turbulence, which can in turn lead to an adverse impact on the performance and fatigue life of wind turbines \cite{nebenfuhr2017prediction}. In recent years, there has been increased focus on the siting of wind turbines within forests. This is in part motivated by the opportunity to reduce the visual and environmental noise footprint of turbines, since the forest canopy and background noise provide a natural screen. It is also driven by need, since around 30\% of the Earth’s land surface is covered by forest. It is thus increasingly relevant to find an effective method to simulate wind field under the influence of forests to help inform the siting at the early design phase of a wind farm.

The effect of forests 
may be considered as a kind of surface roughness which is modelled by introducing a roughness length\cite{lo1990determination}. However, this method is only applicable to short vegetation. The permeability of the foliage should also be taken into account for relatively high forests. Shaw and Schumann \cite{shaw1992large} presented the effect of forests in their simulations as a drag force applied by the trees on the flow and the drag force varies according to the varying leaf area density (LAD). Their method has been accepted by many researchers and has been used to investigate many aspects of canopy flows, for example, coherent structures in canopy flow \cite{nezu2008turburence,dupont2009coherent,finnigan2009turbulence}, the influence of LAD on canopy flow \cite{dupont2008influence} and edge flow due to heterogeneity in forest structure\cite{dupont2008edge}.

%
Large-eddy simulation (LES) has proven to be an useful tool to unravel the engineering problems such as wind farms flow prediction\cite{stevens2014large, stevens2015coupled, stevens2017flow}. 
The atmospheric boundary layer (ABL) can be simulated using conventional finite difference or finite volume methods solvers, for instance, PALM \cite{maronga2015parallelized}, ICON \cite{dipankar2015large}, and MicroHH \cite{van2017microhh}. There are many studies which have endeavored to combine LES with forest models. Shaw and Schumann \cite{shaw1992large} successfully represented a forest as a drag layer and heat sources in their LES. Shaw and Patton \cite{shaw2003canopy} further refined the forest model by including skin friction effect under the LES framework. 

LES is commonly adopted in the Navier-Stokes (NS) equation framework. However, studies have been conducted to involve LES into the framework of lattice Boltzmann models (LBM)\cite{hou1994lattice}. 
In the past decades, LBM has become more widespread on account of its broad applicability across a range of complex fluid dynamics problems ranging from micro-nano scales\cite{Belardinelli15, xue2018effects, xue2020brownian} to macroscopic scales\cite{Filippova2001a, toschi2009lagrangian} under low Mach numbers. 
Moreover, its intrinsic locality, whereby only information from directly adjacent points is required, renders the algorithm efficient for parallelization \cite{Kruger2017}. LBM is thus seen as a good alternative to more common methods based on NS solvers. A recent study combined LBM and LES to simulate ABL flows~\cite{feng2021prolb}. They simulated neutral, stable and convective ABLs over a forest in reference to the cases in~\cite{nebenfuhr2015large}, where LES implemented with a finite volume method was used. Their simulations included the forest effects using a forest model. Furthermore, since the mesh resolution near the ground wall was not sufficient to resolve small-scale turbulent fluctuations, the effect was recovered using the Monin-Obukhov wall model. Their simulations results showed good agreement with the reference data in~\cite{nebenfuhr2015large}.

The present study aims to explore the efficiency of an LBM solver based on a GPU architecture, applied to the simulation of ABL flows with forests. Like other works in the literature, LES is used to ensure the accuracy. 
Various mesh refinement techniques have been proposed for LBM on GPUs, but they suffer from reduced computational efficiency. Another concern is that as lattice points remain in a ‘cubic’ structure, the mesh refinement is generally limited to octree based methods with a scale of 1-2 in each direction. It means that high aspect ratio cells, which are often used in wall-bounded turbulent flows, are generally not possible. To avoid loss of the computational efficiency, meshes with uniform cells can be adopted for ABL simulations on GPUs. In principle, however, a wall function should be used because the near-wall mesh resolution is not sufficient to resolve small scale flow fluctuations. When the forest effects near ground walls are modelled, questions arise about whether the forest drag force is dominant and whether the wall function can be neglected. These are the motivations for the present study.

\section{Methodology}\label{sec:methodology}

\subsection{The lattice Boltzmann method}\label{sec:method-lbm}
The flow is described based on the mesoscale lattice Boltzmann equations. The classical Bhatnagar–Gross–Kroog (BGK) LBM collision operator and D3Q19 model are adopted for the LBM. This model is a three-dimensional lattice model with 19 discretized velocity directions $\bv{c}_i$ ($i=0...Q-1$). We consider $f_{i}(\mathbf{x},t)$ as the discretized particle's probability distribution function on the $i$-th direction of a lattice cell. The lattice cell is located at position $\mathbf{x}$ at time $t$. The LBM governing equation for the fluid, considering collision and forcing, can be written as:

\be
\label{eq:lbe}
f_i(\mathbf{x}+\mathbf{c}_{i}\Delta t,t+\Delta t) =f_i(\mathbf{x}, t) + \Omega\left[f_i(\mathbf{x},t )-f_i^{\mbox{\tiny eq}}(\mathbf{x},t )\right] + \Delta t F_i(\mathbf{x}, t),\\
\ee

where $\Omega$ is a collision kernel~\cite{succi2001lattice, kruger2017lattice}, $\Delta t$ is the marching time step, $F_i$ is the volume force acting on the fluid following the approach from Guo~\cite{guo2002discrete}. 

The collision kernel is the classical BGK collision kernel which has been widely adopted to various applications~\cite{succi2001lattice}. The BGK collision operator fixes the single relaxation time $\tau$ for the colliding process. Thus, the collision operator can be represented as:

\be
\label{eq:bgk}
\Omega\left[f_i(\mathbf{x},t )-f_i^{\mbox{\tiny eq}}(\mathbf{x},t )\right] = \frac{\Delta t}{\tau}\left[f_i(\mathbf{x},t )-f_i^{\mbox{\tiny eq}}(\mathbf{x},t )\right]
\ee

The collision kernel relaxes the distribution function towards the local equilibrium $f_i^{\mbox{\tiny eq}}$:

\be
\label{eq:local_eq}
f_i^{\mbox{\tiny eq}}\left(\textbf{x},t\right) =  \omega_{i}\rho\left(\textbf{x},t\right) \bigg[1+\frac{\textbf{c}_i\cdot\textbf{u}\left(\textbf{x},t\right)}{c_s^2}+\frac{\left[\textbf{c}_i\cdot\textbf{u}\left(\textbf{x},t\right)\right]^2}{2c_s^4}  - \frac{\left[\textbf{u}\left(\textbf{x},t\right)\cdot\textbf{u}\left(\textbf{x},t\right)\right]}{2c_s^2}\bigg] \; ,
\ee

where $\omega_{i}$ is a suitable weight needed to impose the isotropy in the interaction, $\rho\left(\textbf{x},t\right)$ and $\textbf{u}\left(\textbf{x},t\right)$ are the macroscopic hydrodynamic quantities for density and velocity, respectively. The volume force $F_i(\mathbf{x}, t)$ in~\cref{eq:lbe} can be obtained by:

\be
\label{eq: bodyforce}
F_i(\mathbf{x}, t) = (1 - \frac{1}{2\tau}) \omega_i \left [ \frac{\textbf{c}_i - \textbf{u}\left(\textbf{x},t\right)}{c_s^2} + \frac{\textbf{c}_i \cdot \textbf{u}\left(\textbf{x},t\right)}{c_s^4}\textbf{c}_i \right] \mathbf{F},
\ee

where $\mathbf{F}$ is the volume acceleration.

The macro-scale quantities for the density, momentum, and momentum flux tensors can be calculated from the distribution function, the discrete velocity, and the volume force:

\begin{equation}\label{eq:density}
\rho(\mathbf{x}, t) = \sum_{i=0}^{Q-1} f_i(\mathbf{x}, t), \\
\end{equation}

\be
\label{eq:momentum}
\rho(\mathbf{x}, t)\vv(\mathbf{x}, t) = \sum_{i=0}^{Q-1} f_i(\mathbf{x}, t)\mathbf{c}_{i} + \frac{1}{2}\mathbf{F}\Delta t,
\ee

\be
\label{eq:tensor}
\mathbf{\Pi}(\mathbf{x}, t) = \sum_{i=0}^{Q-1} f_i(\mathbf{x}, t)\mathbf{c}_{i}\mathbf{c}_{i},
\ee
where the momentum flux, $\mathbf{\Pi}$, can be presented by the sum of the equilibrium and non-equilibrium parts, $\mathbf{\Pi}(\mathbf{x}, t)  = \mathbf{\Pi}_{\text{eq}} (\mathbf{x}, t)  + \mathbf{\Pi}_{\text{neq}}(\mathbf{x}, t)$.

\subsection{Smagorinsky subgrid-scale model}\label{sec:method-sgs}

The basic idea of LES is not only to resolve relatively large scales, but also to use a subgrid-scale (SGS) model to estimate unresolved small scales. The detailed implementation of the Smagorinsky SGS model can be found in~\cite{hou1994lattice, Koda2015}. Here, we briefly summarize the essential part of this model. The key of the LES modeling of Smagorinsky is to model the effective viscosity $\nu _{\text{total}}$, which can be seen as the sum of the molecular viscosity $\nu_0$ and the turbulent viscosity $\nu_t$:
%

\begin{equation}
\nu _{\text{total}}=\nu _0+\nu_t,
\label{smogrinsky_model}
\end{equation}
where $\nu_t$ is represented by:

\be
\label{eq: nu_t}
\nu_t = C_{smag}\Delta ^2\left |\bar{\mathbf{S}}\right |,
\ee
where $_{smag}$ is the Smagorinsky constant, $\Delta$ represents the filter size, and $\left |\mathbf{\bar{S}} \right |$ is the filtered strain rate tensor, which can be obtained by:

\be
\label{eq: strain}
\left |\bar{\mathbf{S}}\right | = \frac{-\tau_0 \rho c \Delta x + \sqrt{(\tau_0 \rho c \Delta x)^2 + 18\sqrt{2}\rho C_{smag} \delta^2 Q^{1/2}}}{6 \rho C \delta^2},
\ee
where $\tau _0$ is the original relaxation time from the input, $c = \Delta x / \Delta t$, and $Q^{1/2}$ can be written as:

\be
Q^{1/2} = \sqrt{\mathbf{\Pi}_{\text{neq}}:\mathbf{\Pi}_{\text{neq}}},
\ee
where $\mathbf{\Pi}_{\text{neq}}$ is the non-equilibrium part of the momentum flux tensor $\mathbf{\Pi}$ shown in~\cref{eq:tensor}.
Following \cite{Koda2015}, we can obtain the total relaxation time $\tau _{\text{total}}$, which is written as:

\begin{equation}
    \tau _{\text{total}}=\frac{\tau _0}{2}+\frac{\sqrt{(\tau_0\rho c)^2 + 18\sqrt{2}C_{smag} Q^{1/2}}}{2\rho c}.
\end{equation}

Finally, we replace $\tau$ in~\cref{eq:bgk} with $\tau _{\text{total}}$ and obtain the Smagorinsky SGS collision kernel.

\subsection{Forest model near the wall}\label{sec:forest_model}

We model the effects of the forest by introducing an additional drag force near the wall. The drag force is modelled with the help of the forest drag coefficient $C_D$, the LAD $a_f$ and the local wind velocity $\boldsymbol{u}= (u, v, w)$ \cite{shaw1992large}. Thus, the drag force of the forest, which is added in the streamwise direction, can be written as:

\begin{equation}
F_f(x,y,z,t)=-C_D a_f(y) \left | \boldsymbol{u}(x,y,z,t )  \right | u(x,y,z,t ),
\label{eq:forestForce}
\end{equation}
where $y$ refers to the vertical distance to the wall and $u (x, y, z)$ is the streamwise velocity. The forest drag coefficient $C_D$ is usually taken between $0.15$ and $0.2$ \cite{dupont2008influence,}. In this study, $C_D=0.15$. 

According to \cite{lalic2004empirical}, the LAD distribution $a_f(y)$ can be described by the function:

\begin{equation}
a_f(y)={a_f}_m\left(\frac{h-y_{m}}{h-y}\right)^{n} exp{ \left[n\left(1-\frac{h-y_{m}}{h-y}\right)\right]}
\label{af_theory}
\end{equation}
where
\begin{equation}
n=\left\{\begin{array}{ll}
6 & 0 \leq y<y_{m} \\
\frac{1}{2} & y_{m} \leq y \leq h
\end{array}\right.
\end{equation}

The parameter $h$ is the total height of the forest, and $y_m$ is the location where $a_f(y)$ gets its maximum value ${a_f}_m$.

\section{Numerical Implementation}\label{sec:implementation}

The simulations presented in this paper are performed using the open source CFD software GASCANS, which implements LBM on multiple graphic processing units (GPUs) using CUDA and C++. The main aim of GASCANS is to provide fast and accurate simulation of turbulent flows over complex geometries. To further increase its usability, it incorporates a synthetic eddy method (SEM) to generate turbulent velocity fluctuations from mean flow data and  GASCANS can be coupled at run time with another CFD software \cite{Camps2021}. It also implements immersed boundary method (IBM) and a structural solver for the simulation of filaments submerged in fluid.  

The flow solver implements the BGK collision scheme, the LES Smagorinsky turbulence model and forcing scheme described in section \ref{sec:methodology}. The algorithm follows a merged stream-collision time loop and stores the particle distribution functions using a double-population scheme \cite{Latt2021}. The simulation data is stored in a mesh formed by equally sized cubic cells using a structure of arrays (SOA) pattern.  

\begin{figure}[H]
\centering
\includegraphics[width=0.8\linewidth]{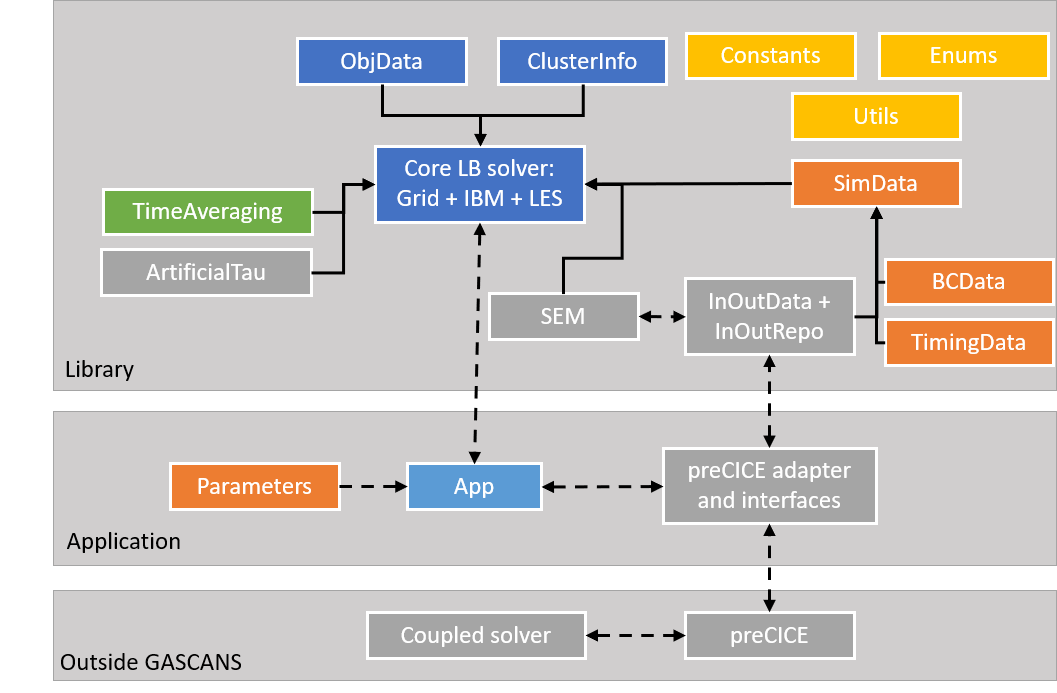}
\caption{Illustration of GASCANS architecture and relationship between its components. The solid arrows indicate aggregation (i e. the classes at the beginning of the arrow are part of the class that the arrow points to) and the dashed arrows indicate primary communication. The isolated boxes represent definitions and utilities accessible by the core LB solver.}
\label{GASCANSClasses}
\end{figure}

~\cref{GASCANSClasses} shows the GASCANS architecture and interactions between its components; the features not used for this paper are greyed out. The code is divided into two sections: the Application contains the user interface and inputs, while the Library contains the core software. This structure allows for the Application to be reworked or substituted to suit the user's needs without modifying the core of GASCANS. The user inputs are through the Parameters class, and the terrain and forest geometry is read from point cloud files. See \cite{Camps2021} for a detailed explanation of the GASCANS functionality and structure.  

The point cloud file contains a list of coordinates covering the forest's volume. GASCANS reads the file and assigns a code to each cell that contains one or more forest coordinates. The code for each forest cell contains the following information: 

\begin{itemize}
    \item $h$: height of the forest, calculated as the bounding box of the forest object in the vertical direction. 
    \item $y$: distance in the vertical direction between the lower bounding box of the forest 
    object and the current cell.
    \item $y_m$: distance in the vertical direction between the lower bounding box of the forest and the maximum value of the LAD distribution. 
\end{itemize}

This information is stored into the forest cell's code using a bit mask to reduce the GPU memory required to store the information. Algorithm \ref{alg:forestGASCANS} shows the forest implementation in GASCANS.

\begin{algorithm}[H]
   \caption{ Implementation of forest forcing with varying leaf area density in GASCANS.}
   \label{alg:forestGASCANS}
   \begin{algorithmic}
    \STATE Read the user input data: $C_{D}$, $a_{f_m}$, $y_m$ in the vertical direction. 
    \STATE Read the forest geometry and generate the forest code for each forest cell. 
    \STATE Set the body force for all cells to 0.
    \FOR{time = 0 to time = last time}
        \FORALL{cells}
            \STATE Stream
            \IF{cell $\mathbf{x}$ is forest}
                \STATE Calculate local cell velocity $\mathbf{u}(\mathbf{x}, t)$
                \STATE Calculate forest force~\cref{eq:forestForce}
                \STATE Correct the local cell velocity using the forest force
            \ENDIF
            \STATE Collide using the calculated body force~\cref{eq:lbe}
        \ENDFOR
    \ENDFOR
   \end{algorithmic}
\end{algorithm}


\section{Validation of simulation platform}\label{sec:num-setup}

Channel flow is a typical case of wall-bounded turbulence. The accuracy of the simulation platform, which uses the LES with the Smagorinsky SGS model, is validated based on the database of direct numerical simulation (DNS) at $Re_\tau=180$ in~\cite{Moser1999}. 

\begin{figure}[H]
\centering
\includegraphics[width=.45\textwidth]{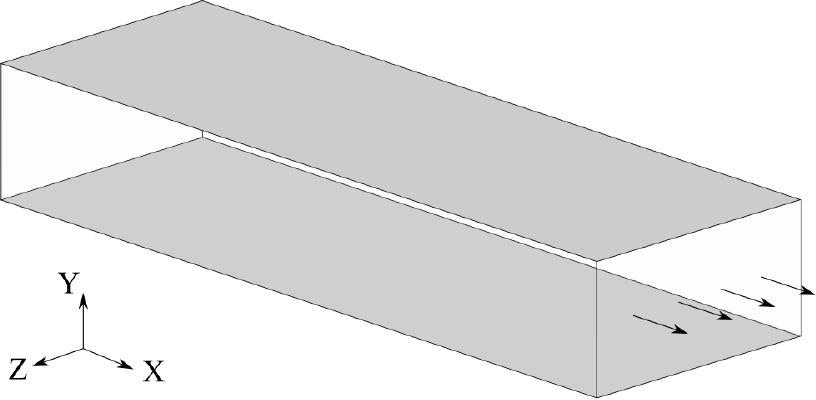}
\caption{The computational domain of the turbulent channel flow. The domain dimensions, normalized with the length scale $H$, are $L_x \times L_y \times L_z = 12 \times 2 \times 4$ along the $x$, $y$ and $z$ axes. The walls are marked in gray, and the arrows indicate the flow direction. }
\label{channel_configuration}
\end{figure}

The computational domain is shown in~\cref{channel_configuration}. The domain dimensions are non-dimensionalized based on the characteristic length $H$. That is, the dimensions are $L_x \times L_y \times L_z = 12 \times 2 \times 4$ in the streamwise, vertical and spanwise directions. The mesh consists of cubic lattice cells. There are $31$  cells per length of $H$. Correspondingly, the mesh contains $372 \times 124 \times 62$ lattice cells. 

The upper and lower walls are set with the no-slip boundary condition, and the remaining boundaries of the computational domain with the periodic boundary condition. 
The flow is driven by a volume body force from~\cref{eq: bodyforce}, which is defined as:   
\begin{equation}
    \mathbf{F}=[\frac{\rho u_{\tau}^2}{H}, 0, 0].
    \label{Body_force}
\end{equation}
The body force acts equivalently as the pressure gradient that balances the wall shear stress in the fully developed channel flow.

The parameters set for the simulation are listed in~\cref{case1_paraview}. Here $C_{smag}$ is the Smagorinsky model constant, as shown in~\cref{eq: nu_t}. The dimensionless density $\rho$ is set to unity in the lattice Boltzmann unit (LBU). The fluid is incompressible. 

\begin{table}[H]
  \centering
  \caption{The simulation parameters for the turbulent channel flow}
    \begin{tabular}{cc|cc}
    \hline
    \hline
    Parameter  & Value & Parameter  & Value \\
    \hline
    $\Delta x$ [m]  & $1/31$    & $L_x$ [m] & 12 \\
    $\Delta t$ [s] & $0.001$ & $L_y$ [m] & 2 \\
    $\nu$ [$m^2$/s]   & $1/3250$  & $L_z$ [m] & 4 \\
    $F$ [N] & 0.002986 & $C_{smag}$     & 0.01 \\
    \hline
    \hline
    \end{tabular}%
  \label{case1_paraview}%
\end{table}%



\begin{figure}[H]
  \centering
    \includegraphics[width=.6\textwidth]{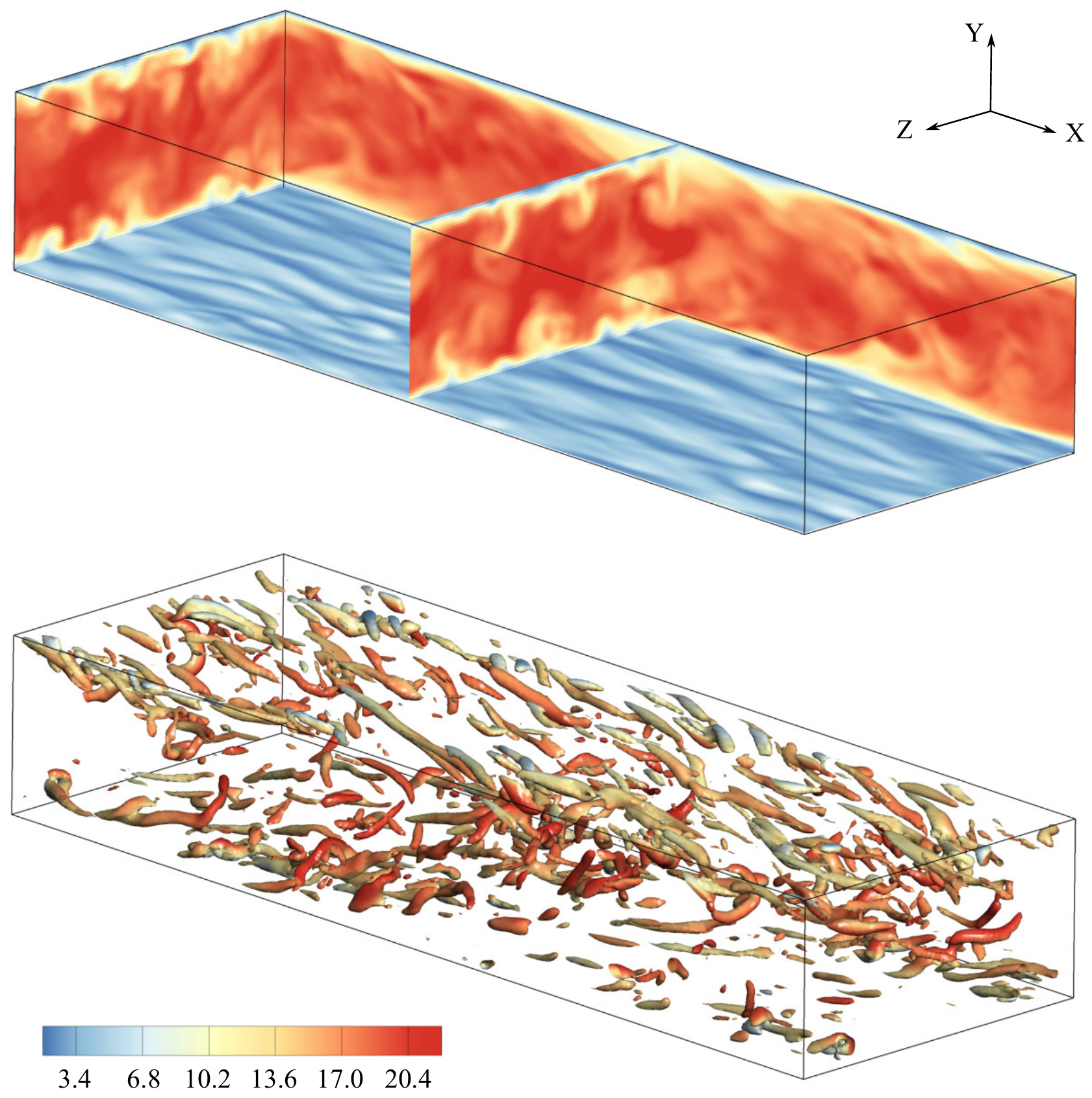}
  \caption{(a) The instantaneous non-dimensional streamwise velocity, $u / u_\tau$, of the turbulent channel flow at $Re_{\tau} = 180$. The bottom plane shown here is located at $y^+ = 9$. (b) Isosurfaces of the Q-criteron of $Q = 0.0002$, which are colored with the streamwise velocity.}
  \label{channel_flow_field_double}
\end{figure}

Figure~\ref{channel_flow_field_double} shows the instantaneous streamwise velocity distribution and the Q-criterion of $Q=0.0002$. 
A plane is placed at the vertical position of \textbf{$y^{+} = 9$} to visualize streamwise velocity contours. Streaks featuring the boundary layer are observed. This is consistent with the previous DNS~\cite{Moser1999}. Plentiful turbulent structures are identified within the flow field based on the Q-criterion. Hairpin vortices exist near the walls, leading to streak contours of the streamwise velocity. 

\begin{figure}[H]
  \centering
    \includegraphics[width=0.85\textwidth]{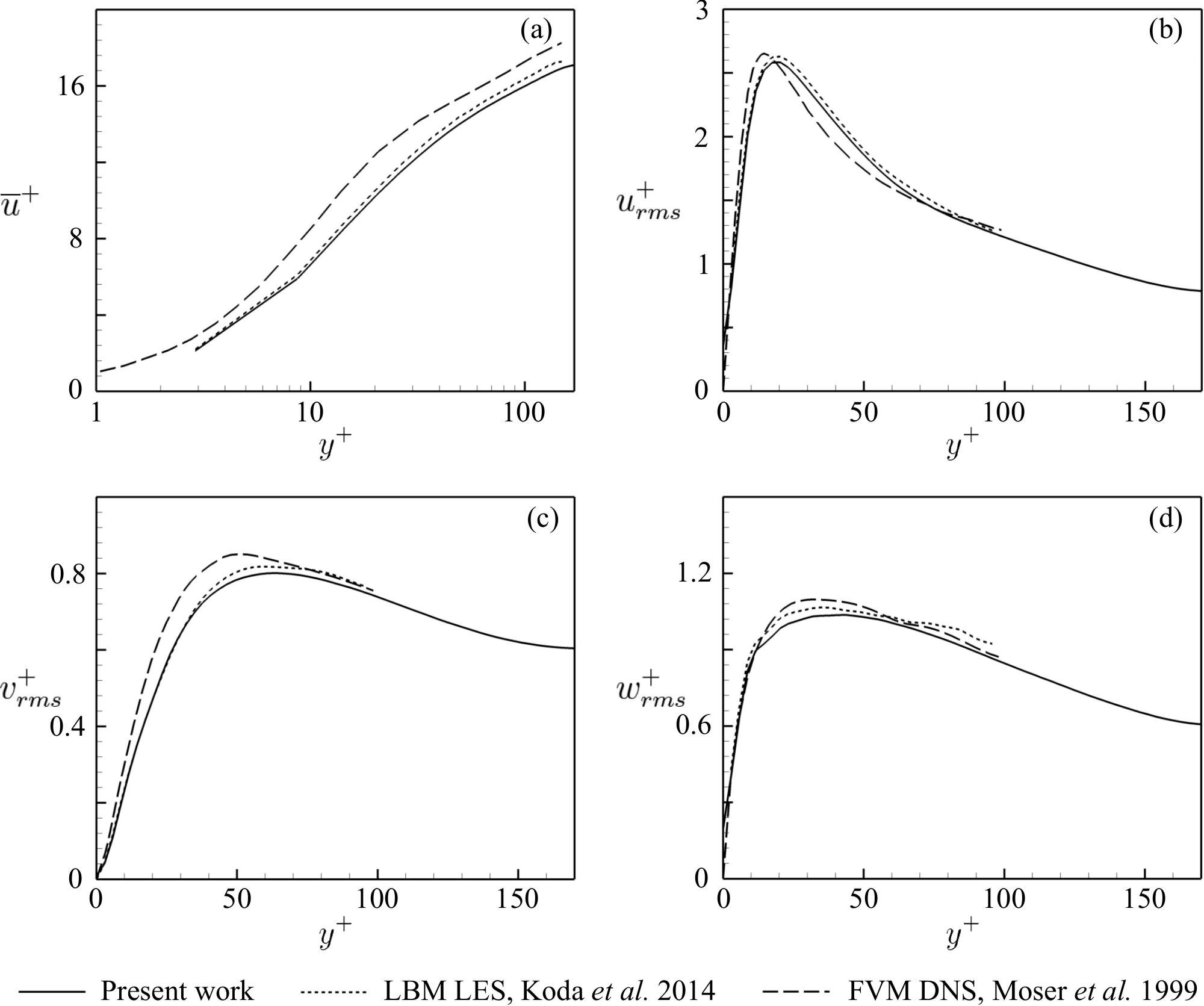}
  \caption{The comparison of the present results and the DNS data computed using LBM of LES~\cite{Koda2015} and a finite volume method of DNS~\cite{Moser1999}. (a) The non-dimensional averaged streamwise velocity $\langle{u}^+\rangle$ as a function of $y^+$. (b-d) The non-dimensional RMS velocity components in the streamwise, wall normal and spanwise directions.
  }
  \label{channel_flow_field_triple}
\end{figure}

As shown in Fig.~\ref{channel_flow_field_triple}, the statistics of the flow quantities from the present LBM LES method are compared to the data computed using the LBM LES reported by Koda et al. \cite{Koda2015} and the finite volume method of DNS by Moser et al. \cite{Moser1999}. 
The non-dimensional time-averaged streamwise velocity is defined as $\langle {u}^+\rangle = \left<u \right>/u_{\tau}$. Here $\left< \cdot \right>$ is the averaging operator, and $u_{\tau}$ denotes the friction velocity. 
The averaging is done in time and then in x-z planes that are parallel to the top and bottom walls.
The non-dimensional room mean square (RMS) velocity components ($u_{rms}^+$, $v_{rms}^+$ and $w_{rms}^+$) are also normalized with the friction velocity $u_{\tau}$. 
The present results are consistent with Koda's LBM LES data, especially in the near wall region. But these results exhibit discrepancies in comparison with the Moser's DNS data. Both LBM LES methods underestimate $\langle {u}^+\rangle$. The reason is that the mass flow rate computed with the coarse meshes in LES is lower than that from DNS. The RMS velocity components from LES and DNS are mainly different below $y^+=100$. It suggests that the LES method does not completely reproduce the subgrid-scale flow structures near the wall. Nonetheless, the largest discrepancy of $\langle{u}^+\rangle$
is less than $5 \%$, and the largest RMS velocity difference less than $10 \%$.  
Therefore, the present simulation platform, which is developed based on GASCANS, show overall correct prediction in comparison with DNS for wall-bounded turbulence. 

\section{Atmospheric boundary layers over forest}

\subsection{Case description}

\begin{figure}[H]
\centering
\includegraphics[width=0.45\linewidth]{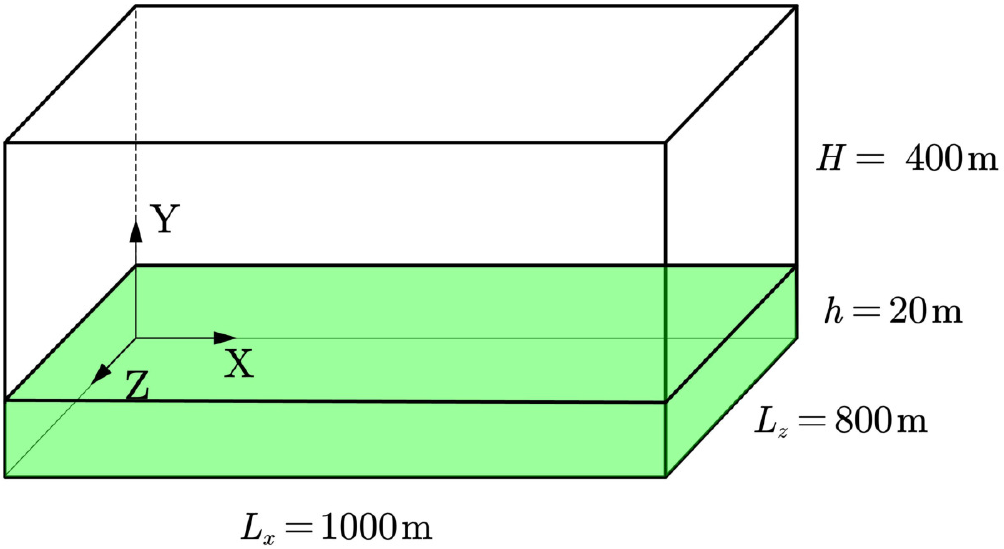}
\caption{The computational domain of the ABL. The green color marks out the zone where a homogeneous forest with a height of $h=20 \; m$ is modelled above the ground. The freestream flow moves along the x-axis.}
\label{forest}
\end{figure}

A classical ABL case reported in~\cite{nebenfuhr2014influence,nebenfuhr2015large}
is studied using the present simulation platform. The computational domain is shown in~\cref{forest}. The physical dimensions are $L_x \times L_y \times L_z = 1000 \times 400 \times 800 \; m^3$. 
To produce the representative forest effects, a forest with a homogeneous distribution over the ground is assumed~\cite{nebenfuhr2014influence}. The height of the forest is $h=20 \; m$. The freestream flow direction is set along the $x$ axis. 
The forest friction velocity is defined as $u_*=\left[ \left< u'v' \right>^2+ \left< v'w' \right>^2 \right] ^{1/4}$ at the location of $y/h = 2$~\cite{nebenfuhr2014influence,nebenfuhr2015large}. 

The flow is driven by a constant non-dimensional body force of 
$2.38\times 10^{3}$ per unit volume on the basis of $u_*$ and $L_x$. 
The force is imposed in the whole domain. At the bottom wall below the forest, a no-slip boundary condition is applied. The inlet and outlet are set with the periodic boundary condition, and the spanwise side boundaries are also set with the same boundary condition. The upper boundary is specified with a forced equilibrium boundary condition. This boundary condition specifies the upper wall to move with the same velocity as the free-stream flow. 
Because the upper wall is sufficiently far away from the bottom, and because only the boundary layer region near the bottom wall where wind turbines are embedded are of interest for the wind energy harvesting, this forced equilibrium boundary condition is suitable for the present ABL case.

\begin{figure}[H]
\centering
\includegraphics[width=0.45\linewidth]{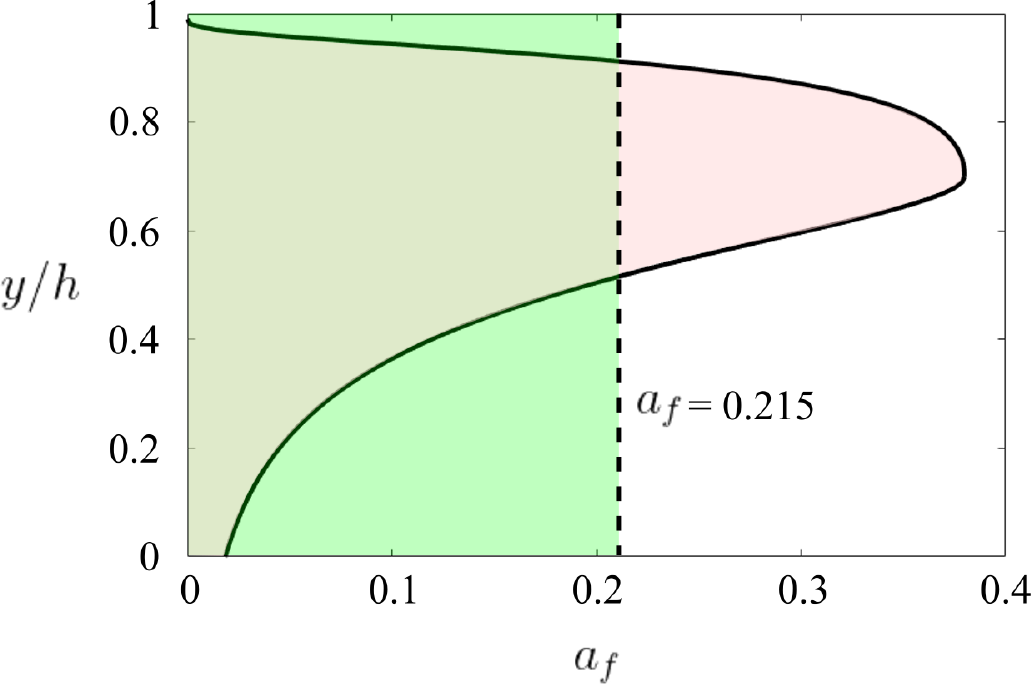}
\caption{The distributions of LAD in two situations such as constant LAD of $a_f=0.215$, indicated by the dash line, and varying LAD with $a_f$ following Eq.(\ref{af_theory}) with $z_m=0.7h$ and ${a_f}_m=0.38$, indicated by the solid curve. The green colors the area of $a_f=0.215$, which is equal to the area of the varying LAD case colored in red. } 
\label{af_constant}
\end{figure}

In reality, the LAD $a_f$ varies with the vertical height of a tree, as the tree crown has dense leaves and the trunk is less obstructing. In the current work, we investigate three distributions of LAD, as listed in Table~\ref{tab:forest_cases}. Case 1 has constant LAD in the vertical direction, indicating that the leaf and obstructions are uniform along a tree. Case 2 has varying LAD, which is more similar to the real situation~\cite{nebenfuhr2014influence,nebenfuhr2015large}. Case 3 has no forest force imposed (i.e., constant LAD of $a_f=0$). However, the LAD distributions of cases 1 and 2 have the same total obstruction, which is defined as $A_f = \int_{0}^{h} a_f (z) dz$, whereas case 3 has no obstruction. The LAD distributions of cases 1 and 2 are illustrated in Fig.~\ref{af_constant}. In case 1, $a_f = 0.215$. In the varying LAD case, $a_f$ is set on the basis of Eq.(\ref{af_theory}) with the coefficients $z_m=0.7h$ and ${a_f}_m=0.38$.

\begin{spacing}{1.0}
\begin{table}[H]
  \begin{center}
  \caption{The simulated ABL cases in terms of the LAD.}
    \begin{tabular}{cccc}
    \hline
    \hline
      & Case 1 & Case 2 & Case 3  \\
    \hline
    $a_f (y)$ & $0.215$ & Eq.~\ref{af_theory} ($z_m=0.7 h$, ${a_f}_m=0.38$) & $0.0$ (i.e., no forest modelling) \\ 
    \hline
    \hline
    \label{tab:forest_cases}%
    \end{tabular} 
  \end{center}
\end{table}%
\end{spacing}

To develop the turbulence, a set of small boxes are set near the inlet in the initial stage of the simulations. The non-dimensional box size is $20\times20\times20$. Once the turbulence is developed, the boxes are removed from the computational domain. The flow remains in the turbulent status afterwards. 

\subsection{Mesh independence study}

The mesh quality for the simulation accuracy is examined with the meshes with coarse, medium and fine cell size. The constant LAD of $a_f=0.215$ is adopted for the forest modelling. The information of the meshes are presented in Table~\ref{tab:mesh_cost}. The coarse mesh has the number of cells in the three dimensions as $N_x \times N_y \times N_z = 200 \times 80 \times 160$, resulting in $2.6 \times 10^6$ cells in total. We define the refinement ratio as the ratio of a refined cell size to the reference cell size. In reference to the coarse mesh, the refinement ratios of the medium and fine meshes are $0.8$ and $1/1.75 \approx 0.57$. 
%

The computational time and memory costs relevant to the meshes are also reported in Table~\ref{tab:mesh_cost}. The GPU, GeForce RTX 2070 SUPER, with a total memory of $8$ GB is used for the computations. 
The coarse mesh simulation consumes the computing time of approximately $8.5$ hours per $1 \times 10^6$ time steps and the memory of nearly $1.44$ GB. For the medium mesh simulation, the computing time is $1.65$ times larger than the coarse mesh, and the memory cost is $1.5$ times larger. The fine mesh simulation is $4.59$ times faster and $4.5$ times more memory consumption. The increasing ratios of the computing time are comparable with those of the memory costs. On the other hand, the total number of cells in the medium and fine meshes increase with ratios of $1.95$ and $5.36$, which are larger than the ratios of the computing time and memory costs. It suggests that the increased mesh size improves the computational efficiency. The reason is that there are modules responsible for processing and storing solution data in the present simulation platform. The additional computational and memory costs from these modules are independent on the mesh size, so they are more weighted in the coarse mesh case, as compared with the other refined mesh cases. 
Moreover, the GPU solves the LBM equations on all cells at the same time. An increased number of cells has less impact on the computing time if the former coarse mesh does not fill all capacity of the GPU.

\begin{spacing}{1.0}
\begin{table}[H]
  \begin{center}
  \caption{The meshes and respective computational costs in the mesh independence study.}
    \begin{tabular}{llllll}
    \hline
    \hline
    Meshes & $N_x \times N_y \times N_z$ $^\S$ & $N_{tot}$ $^\P$ & $\Delta t$ $^\dagger$  & $T_c$ (hours) $^\ddagger$  & Memory (GB)  \\
    \hline
    Coarse & $200 \times 80 \times 160$ & $2.6 \times 10^6$ & $0.3$   & $8.5$  & $1.44$  \\
    Medium & $250 \times 100 \times 200$ & $2.6  \times 10^6 \cdot 1.95$ & $0.2$   & $8.5 \cdot 1.65$   & $1.44 \cdot 1.5$ \\
    Fine & $350 \times 140 \times 280$ & $2.6  \times 10^6 \cdot 5.36$ & $0.1$   &  $8.5 \cdot 4.59$   & $1.44 \cdot 4.5$  \\
    \hline
    \hline
    \label{tab:mesh_cost}%
    \end{tabular} 
    \vspace{-.5cm}
   \begin{tablenotes}
    \small
    \item $\S$ The number of cells in the three dimensions. \\
    $\P$ The total number of cells. \\ 
    $\dagger$ The non-dimensional time interval of each time step. \\
    $\ddagger$ The computing time per $1 \times 10^6$ time steps.  
  \end{tablenotes}
\end{center}
\end{table}%
\end{spacing}

\begin{figure}[H]
    \centering
    \includegraphics[width=.6\linewidth]{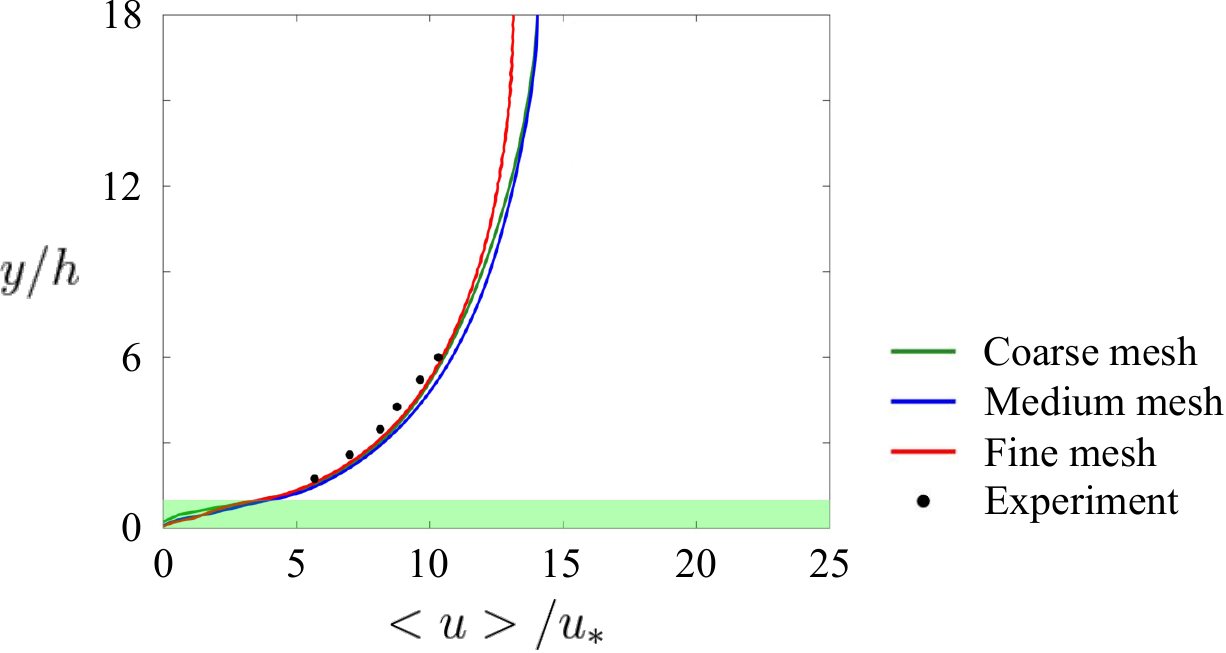}
    \caption{The averaged streamwise velocity obtained from the present simulations with the constant LAD and from the experiments by Bergström et al.~\cite{bergstrom2013wind}}
    \label{fig:mesh_uavg}
\end{figure}

The normalized averaged streamwise velocity along the wall normal direction for the simulations of the three meshes is shown in Fig.~\ref{fig:mesh_uavg}. 
The averaging is done in time and then in $x-z$ planes that are parallel to the bottom wall. 
Hereafter, the averaged quantities reported in this study are obtained using the same averaging method. 
As compared to the experimental data from Bergstr\"{o}m et al.~\cite{bergstrom2013wind}, the fine mesh simulation shows the best agreement. In the range of the experimental data, the coarse mesh simulation gives better prediction than the medium mesh simulation. But these two simulations become similar above $y/h=14$, which is far from the bottom wall and forest. The profile of the coarse mesh simulation rises more slowly for $y/h > 2$. It implies that there are less turbulent vortices resolved by the coarse mesh. 

\begin{figure}[H]
    \centering
    \includegraphics[width=0.8\linewidth]{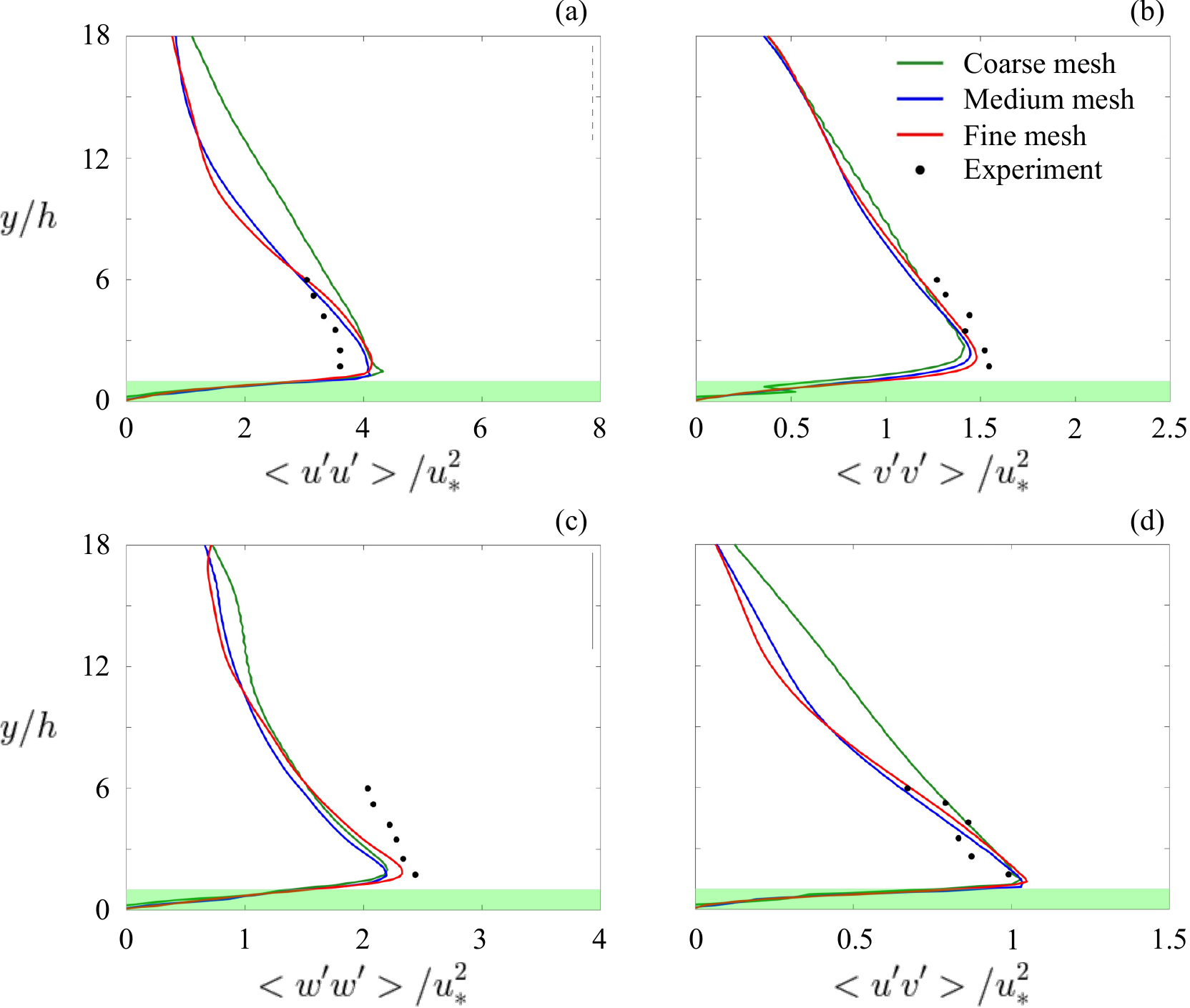}
    \caption{The normalized Reynolds stress tensor components computed with the three meshes and reported in the experiments by  Bergström et al.~\cite{bergstrom2013wind} Here the constant LAD is set in the forest modelling.}
\label{fig:mesh_str}
\end{figure}

The Reynolds stress tensor components, $\left< u'u' \right>$, $\left< v'v' \right>$, $\left< w'w' \right>$ and $\left< u'v' \right>$, normalized with the averaged wall friction velocity, $u_*$, are shown in Fig.~\ref{fig:mesh_str}. 
It is observed that the coarse mesh leads to the worst prediction of all components in reference to the experimental data from Bergström et al.~\cite{bergstrom2013wind}, whereas the prediction of the fine mesh is most consistent. Therefore, a general trend is that the mesh refinement improves the simulation accuracy. 

For the streamwise component $\left< u'u' \right> / u_*^2$ in Fig.~\ref{fig:mesh_str}, the slope of the profile from the coarse mesh simulation is more straight than the other meshes. This is a typical phenomenon when the turbulent flow is less developed. As a consequence, velocity fluctuations owing to turbulence vortices are insufficient to form a steeper profile. This effect is also found for the wall normal component $\left< v'v' \right> / v_*^2$ and the spanwise component $\left< w'w' \right> / w_*^2$, but it is not so significant as $\left< u'u' \right> / u_*^2$. Since streamwise velocity fluctuations have larger magnitudes than spanwise and wall normal velocity fluctuations, the streamwise velocity ones are more influential in the formation of the Reynolds stress component $\left< u'v' \right>$. As seen in subfigure~\ref{fig:mesh_str}d, the profiles of $\left< u'v' \right> / u_*^2$ show similar trends to $\left< u'u' \right> / u_*^2$.

The fine mesh results in overall the best prediction, as indicated in Figs.~\ref{fig:mesh_uavg} and \ref{fig:mesh_str}. Based on Tabl~\ref{tab:mesh_cost}, the averaged first-layer cell height over the wall is $\left< \Delta y+ \right> =165$. Moreover, the results of the medium and fine meshes are nearly the same. This converged prediction means that the fine mesh resolution is sufficient to simulate the current ABL case. In addition, the computational time and memory costs of this mesh are lowest, which is found in Table~\ref{tab:mesh_cost}. The fine mesh is therefore adopted in the following simulations and analysis. 

Based on Tabl~\ref{tab:mesh_cost}, the averaged first-layer cell height over the wall is $\left< \Delta y+ \right> =165$. This infers that the wall modelling should be used in principle. However, the practice in the current simulations show that good agreement is achieved without using a wall function. The reason is that the forest forces are much larger than the wall friction. This effect will be discussed in the subsequent section on the forest modelling. 

\subsection{Effects of forest modelling}

To investigate the effects of the forest modelling, the simulation cases with different LAD are defined in Table~\ref{tab:forest_cases}. Case 1 (the modelled forest with the constant LAD) has been discussed in the preceding section, and it is found producing consistent prediction in comparison to the experiments. 

\begin{figure}[H]
\centering
    \includegraphics[width=0.9\linewidth]{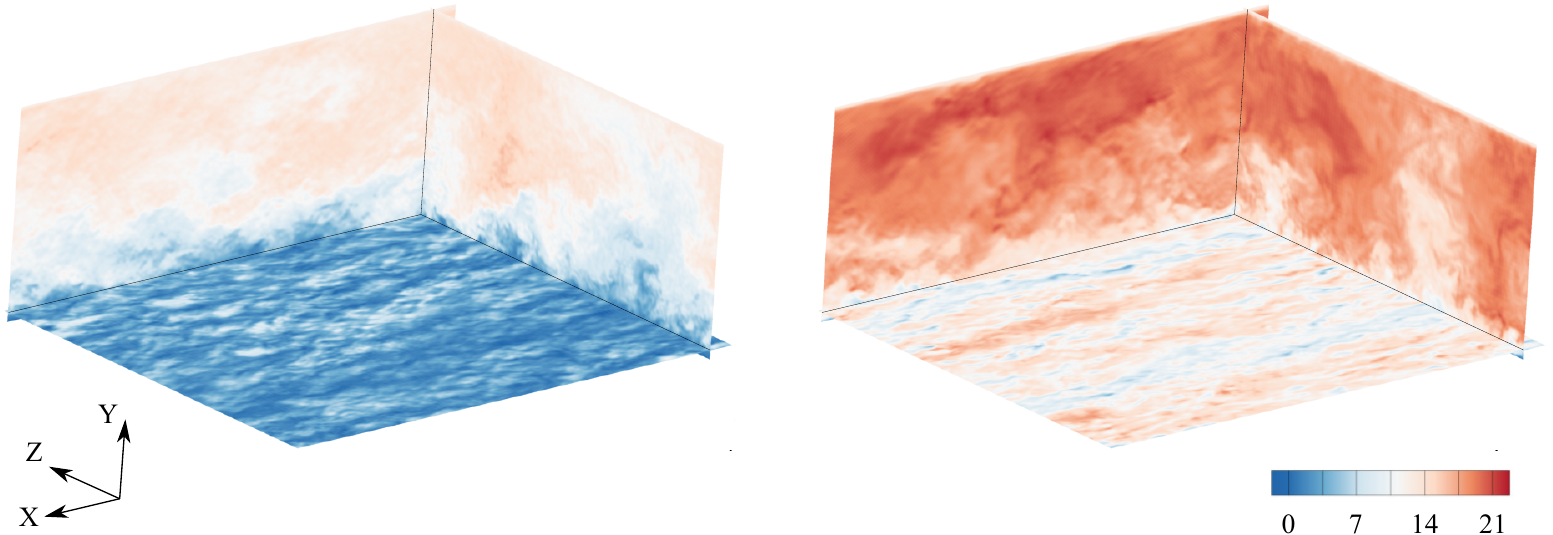}
\caption{A snapshot of the instantaneous normalized streamwise velocity $u / u_*$. (a) The simulation with the forest modelling of the varying LAD; (b) the simulation with no modelled forest. The bottom surface visualized here is located at $y/h=0.5$, which is the middle of the forest zone.}
\label{fig:cases23_u}
\end{figure}

For case 2 (the modelled forest with the varying LAD) and case 3 (zero LAD, i.e., no modelled forest), the instantaneous normalized streamwise velocity $u / u_*$ are shown in Fig.~\ref{fig:cases23_u}. Case 2 exhibits generally much smaller velocity fluctuations than case 3 in the whole computational domain. 
Fluctuations are visualized at a $x-z$ plane of $y/h=0.5$, which is the center position of the modelled forest zone. Velocity magnitudes in this plane in both cases are smaller than those in the region far from the bottom wall. As case 2 includes the drag forces from the forest modelling and from the wall friction, the small fluctuations within the forest zone are explainable for this case. There are no forest drag forces imposed in case 3. In other words, case 3 does not have a forest zone. Its near-wall flow is only dragged by the wall friction. 
Therefore, the forest dragging effect is dominant in reducing the velocity of the ABL flow, as compared with the wall friction.
It is observed in case 2 that the velocity quickly increases to larger values outside the forest zone. This also qualitatively indicates that the significant dragging effect from the modelled forest. The same finding was reported in~\cite{nebenfuhr2014influence,nebenfuhr2015large}.

\begin{figure}[H]
\centering
    \includegraphics[width=0.7\linewidth]{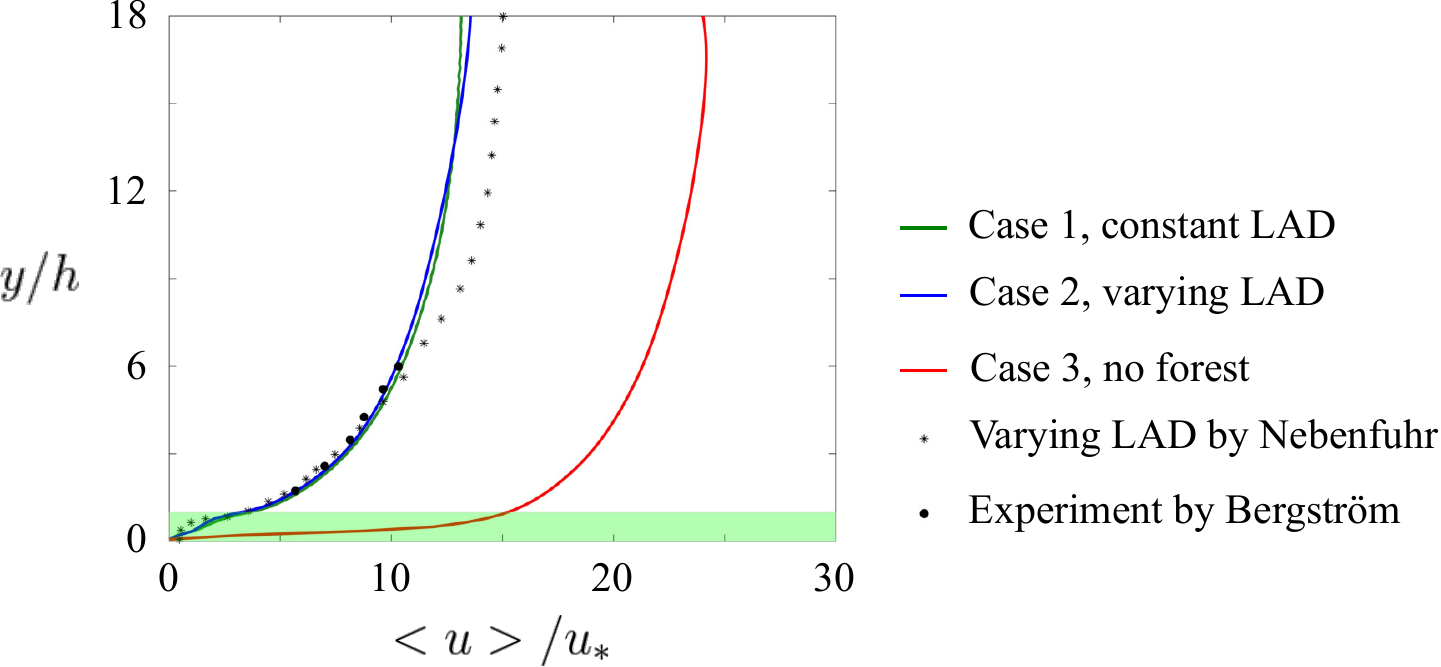}
\caption{The averaged streamwise velocity in the present simulations with different forest modelling settings, as well as from the simulations using the FVM LES by Nebenf\"{u}hr et al.~\cite{nebenfuhr2014influence,nebenfuhr2015large} and the experiments by Bergstr\"{o}m et al.~\cite{bergstrom2013wind}}
\label{fig:uplus_all}
\end{figure}

The vertical profiles of the averaged streamwise velocity $\left< u \right> / u_*$ in all simulated cases listed in \ref{tab:forest_cases}, as well as the FVM LES with the same varying LAD setting as case 2 in~\cite{nebenfuhr2014influence,nebenfuhr2015large} and the experiment data in~\cite{bergstrom2013wind}, are plotted in Fig.~\ref{fig:uplus_all}. 
In the near ground zone of $y/h <6$ where large part of the boundary layer is located, cases 1 and 2 exhibit good agreement with the experimental data. The FVM LES also agrees with the experiment, while discrepancies are observed for $y/h$ above approximately $5$. The 
The mesh for the FVM LES contains $10$ cells in the forest zone. It has been demonstrated that the mesh resolution is sufficient to resolve the profile of the varying LAD illustrated in Fig.~\ref{af_constant} \cite{nebenfuhr2014influence,nebenfuhr2015large}. There are $7$ cells in the fine mesh for the current simulations, which also resolves the same varying LAD profile. The comparison of cases 2 and 3 indicates that the constant and varying LADs lead to slight differences in the prediction of the mean velocity in the forest zone and most of the zone above the forest. 

In case 3 that has no forest, the boundary layer development is completely different from cases 1 and 2 considering the forest effects. The velocity gradient in the forest zone in case 3 is much larger than the other cases. A similar phenomenon was also reported in~\cite{nebenfuhr2014influence,nebenfuhr2015large}. The reason is that the forest model introduces significant drag forces into the flow. Based on Eq.~\ref{eq:forestForce}, the forest drag force is independent from the wall friction. Its vertical distribution is not governed by the classical boundary layer statistics, which can be resumed using a wall function. Although the wall function is ignored, only considering the forest model is still able to predict the mean velocity profile matching the experiments. This is explainable because the forest force overwhelms wall friction. 

\begin{figure}[H]
\centering
    \includegraphics[width=0.8\linewidth]{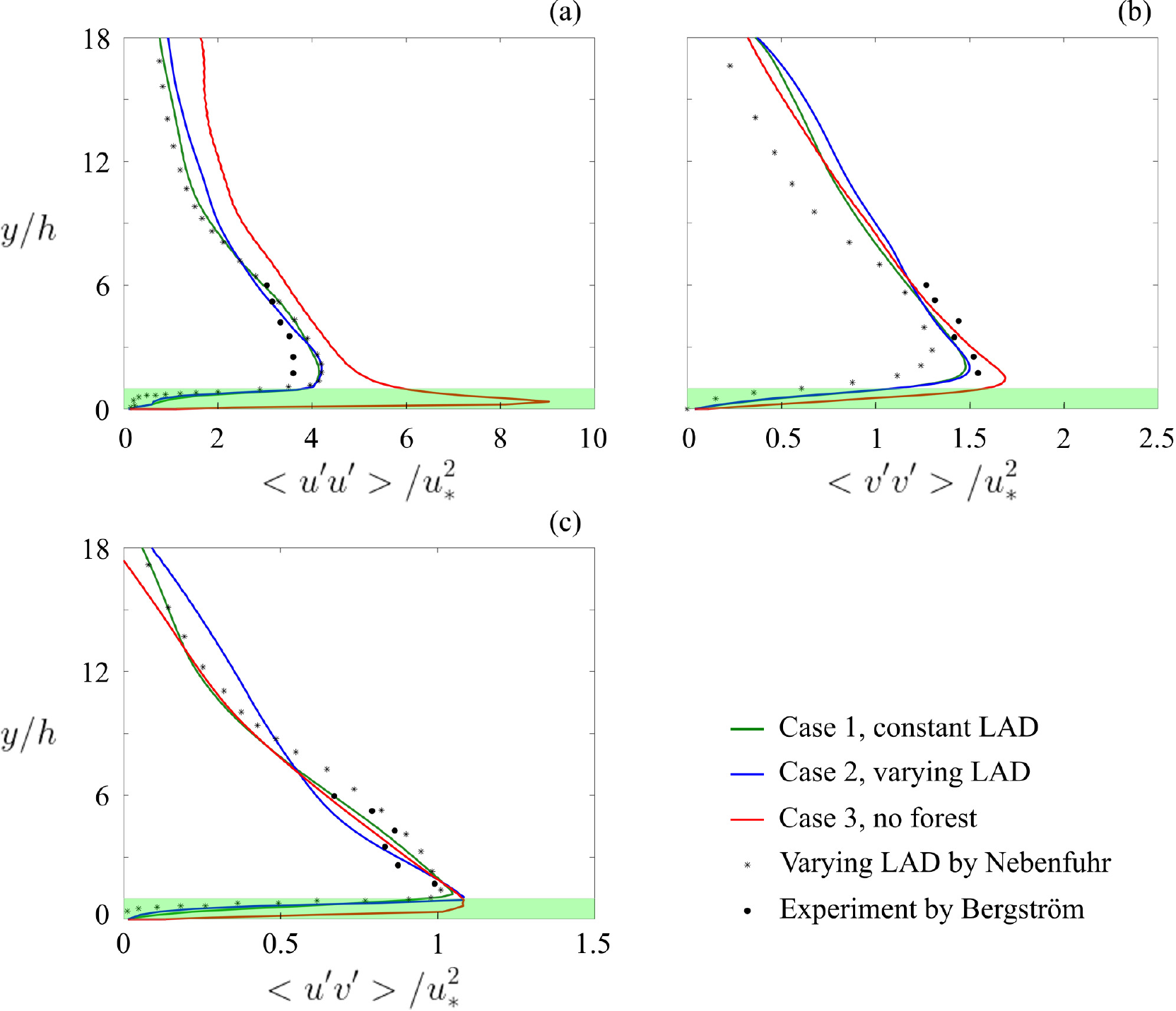}
\caption{The stress tensor components in the present simulations with different forest modelling settings, as well as from the simulations using the FVM LES by Nebenf\"{u}hr et al.~\cite{nebenfuhr2014influence,nebenfuhr2015large} and the experiments by Bergstr\"{o}m et al.~\cite{bergstrom2013wind}.}
\label{fig:forest_str}
\end{figure}

The vertical profiles of the normalized stress tensor components are displayed in Fig.~\ref{fig:forest_str}. In the simulations set with the forest model (cases 1 and 2, as well as the FVM LES in~\cite{nebenfuhr2014influence,nebenfuhr2015large}), the predicted trends of $\left< u' u' \right>$ are consistent. But a mismatch between these methods and the experimental data is observed in the near forest zone of $2 < y/h < 4$. Cases 1 and 2 also show $\left< v' v' \right>$ and $\left< u' v' \right>$ comparable with the experimental data. The different LAD settings of cases 1 and 2 lead to discrepancies between their results. But the results are approximately identical at the first cell height. Recalling Eq.~\ref{eq:forestForce}, the forest drag force modelled is proportional to the local kinetic energy. Since the velocity at the wall is zero based on the no-slip boundary condition, the local forest drag force diminishes there. The small velocity near the wall leads to the negligible force, although the constant LAD is much larger than the varying LAD near the wall (see Fig.~\ref{af_constant}). As the majority difference of the two LAD settings are defined near the wall for $y/h$ less than $0.4$, their influence on the drag force production is limited owing to the small local velocity. As a consequence, the profiles of the tensor statistics simulated with the two LAD settings are similar. Nevertheless, the present study only demonstrates the forest effect modelled with the LADs shown Fig.~\ref{af_constant}. If the LAD definition is changed, the forest effects could change and, thus, should be thoroughly examined. 

As the forest drag force is not modelled, case 3 gives different prediction in comparison with cases 1 and 2, in particular, for $\left< u' u' \right> / u_*^2$ displayed in Fig.~\ref{fig:forest_str}. Obvious differences are identified for all stress tensor components within the forest zone, and $\left< u' u' \right> / u_*^2$ also shows inconsistency at vertical locations far from the forest zone. Therefore, the mismatch conveys that the modelled forest drag force play a critical role in predicting the statistics of flow fluctuations in the boundary layer, in addition to its importance for the mean flow statistics that is shown in Fig.~\ref{fig:uplus_all}.

\section{Conclusions}\label{sec:conclusions}

A GPU-based simulation platform using an LBM method for LES has been proven to be suitable for computing ABL flows with the forest presence, which are of interest in the site selection of wind turbine farms. 
The forest effects are modelled by defining a volume drag force in relation to the local velocity and LAD. The way of implementing the force for LBM is presented in detail. 

To refrain from influencing the GPU computational efficiency, the meshes are not locally refined. Uniform-sized lattices are generated in the whole computational domain. This leads to large $\Delta y+$ near the wall, which in principle requires a wall function. However, we introduce the approximations near the ground wall that the wall function is not taken into account, and that the forest effects are considered. These approximations are made because the forest drag force overwhelms the the wall friction. This is validated based on the qualitative and quantitative analysis of the results. The statistics of the mean flow velocity and stress tensor components are well consistent with the previous FVM-LES simulations and experiments. The mesh independence study illustrates that the modelled forest zone is resolved with $7$ cells.

One advantage of the near-wall approximations is that the LBM program is compatible with the GPU architecture, so the GPU speedup is preserved. Moreover, the wall treatment is ignored to reduce the computational cost. 
The platform extends the capabilities of the open-source simulation tool GASCANS developed at the University of Manchester. 

The forest effects are examined for a classic vertically varying LAD distribution, which was simulated in previous studies using the FVM-LES, and a constant distribution possessing the same total area as the varying one. The results between these two LADs differ only slightly. Nevertheless, it would be interesting to investigate the effects of the LAD vertical profile on the characteristics of ABL flows. In the present study, the standard Smagorinsky model is used to model the SGS turbulence. A future work is to implement other SGS models.

\section{ACKNOWLEDGEMENTS}
The authors would like to acknowledge the funding from Chalmers Area of Advance Transport. The computations and data handling were enabled by resources provided by the Swedish National Infrastructure for Computing (SNIC), partially funded by the Swedish Research Council through grant agreement no. 2018-05973. 


\bibliography{prex}

\end{document}